\documentclass[twocolumn,prd,aps,groupedaddress,showpacs,nofootinbib]{revtex4}

\usepackage{amsmath}
\usepackage{amssymb}
\usepackage{latexsym}
\usepackage{graphicx}
\usepackage{hyperref}
\usepackage{mathrsfs}
\usepackage{color}
\usepackage{bm}
\usepackage[caption=false]{subfig}

\begin{document}

\title{Null test of the cosmic curvature using $H(z)$ and supernovae data }
\author{Rong-Gen Cai$^1$}
\email{cairg@itp.ac.cn}

\author{Zong-Kuan Guo$^1$}
\email{guozk@itp.ac.cn}
\author{Tao Yang$^1$}
\email{yangtao@itp.ac.cn}

\affiliation{\it $^1$State Key Laboratory of Theoretical Physics, Institute of
Theoretical Physics, Chinese Academy of Sciences, P.O. Box 2735,
Beijing 100190, China }


\begin{abstract}
 We introduce a model-independent approach to the null test of the cosmic curvature which is geometrically related to the Hubble parameter $H(z)$ and luminosity distance $d_L(z)$. Combining the independent observations of $H(z)$ and $d_L(z)$, we use the model-independent smoothing technique, Gaussian processes, to reconstruct them and determine the cosmic curvature $\Omega_K^{(0)}$ in the null test relation. The null test is totally geometrical and does not assume any cosmological model. We show that the cosmic curvature $\Omega_K^{(0)}=0$ is consistent with current observational data sets, falling within the $1\sigma$ limit. To demonstrate the effect on the precision of the null test, we produce a series of simulated data of the models with different $\Omega_K^{(0)}$. Future observations in better quality can provide a greater improvement to constrain or refute the flat universe with $\Omega_K^{(0)}=0$.
\end{abstract}
\maketitle
\section{Introduction \label{sec:introduction}}

Whether the space of our Universe is open, flat, or closed is one of the most fundamental problems in modern cosmology. The spatial topology of the Universe is also closely related to other important problems such as the evolution of the Universe, the property of the dark energy, etc. The effect of allowing nonzero curvature on dark energy models has been studied in many papers, see, for example, Refs.~\cite{Ichikawa:2005nb,Ichikawa:2006qb,Clarkson:2007bc,Gong:2006gs}. And the detection of a significant deviation from $\Omega_K^{(0)}=0$ would have profound consequences for inflation models and fundamental physics. A lot of attention has been attracted to this issue~\cite{Eisenstein:2005su,Tegmark:2006az,Zhao:2006qg,Wright:2007vr}. The $\Lambda$CDM model is consistent with all of the data and a flat universe is preferred even in recent Planck 2015 results~\cite{Ade:2015xua}.

However, almost all of the papers studying and constraining the cosmic curvature assume  some specific models for dark energy such as the equation of state $w(z)$. These are all model-dependent and indirect methods. Note the fact that there is some degeneracy between the spatial curvature and the equation of state of dark energy in these studies. It would be better to detect the space curvature of the Universe by geometrical and model-independent methods. In this paper, we use a geometrical relation among the cosmic curvature $\Omega_K^{(0)}$, Hubble parameter $H(z)$, and luminosity distance $d_L(z)$, by combining the Hubble rate and luminosity distance, we are able to directly determine and test whether the cosmic curvature deviates from 0~\cite{Clarkson:2007bc,Shafieloo:2009hi,Li:2014yza}. For this goal, we should focus on two independent observations that directly give $H(z)$ and $d_L(z)$, respectively. For $H(z)$ data, it can be derived from differential ages of galaxies [``cosmic chronometer (CC) "] and from the radial baryon acoustic oscillation (BAO) scale in the galaxy distribution. As for $d_L(z)$, we use the SNIa Union 2.1 data sets. We use the model-independent method Gaussian processes (GP) for smoothing the observational data. The advantage of these methods is that they are all model-independent; hence, we need not assume any models involving dark energy and gravity theory. They are purely geometrical and are constrained directly by observational data. However, the precision of the null test is also limited by the quality of observational data. If we want to detect a tiny cosmic curvature more precisely, a better quality of the observational data sets are also required, which will be also discussed in this paper.

This paper is organized as follows. In Sec.~\ref{sec:tb}, we introduce the theoretical method for the null test of the cosmic curvature. In Sec.~\ref{sec:nulltest}, we first give a brief  introduction of Gaussian processes, and then apply GP method to the null test of the cosmic curvature using two independent data sets: CC+BAO, Union 2.1. Further it is followed by a series of simulated data tests. We give discussions and conclusions in Sec.~\ref{sec:discussion}.

\section{ Theoretical Method \label{sec:tb}}

In a FLRW universe, the luminosity distance $d_L$ can be expressed as
\begin{equation}
{d_L} = \frac{{c(1 + z)}}{{{H_0}\sqrt {\Omega_K^{(0)}} }}\sinh (\sqrt {\Omega_K^{(0)}} \int_0^z {\frac{{d\tilde z}}{{E(\tilde z)}})},
\label{equa:dl}
\end{equation}
where $E(z)\equiv H(z)/H_0$, $\Omega_K^{(0)}\equiv -K {c^2}/(a_0 H_0)^2$, and $K = +1,-1,0$ corresponds to a closed, open, and flat universe, respectively.

Differentiating Eq.~(\ref{equa:dl}) and writing $D(z)=(H_0/c)(1+z)^{-1} d_L(z)$ as the normalized comoving distance, we can obtain
\begin{equation}
\Omega_K^{(0)} = \frac{{{E^2}(z)D{'^2}(z) - 1}}{{{D^2}(z)}}.
\label{equa:ok}
\end{equation}
We can see that the cosmic curvature $\Omega_K^{(0)}$ can be directly determined by using the Hubble parameter and luminosity distance from Eq.~(\ref{equa:ok}). Thus, the null test of $\Omega_K^{(0)}$ is straightforward. Note the fact that $D(0)=0$ will bring about a singularity at $z=0$. Therefore for more succinct, we transform Eq.~(\ref{equa:ok}) to
\begin{equation}
\frac{{\Omega_K^{(0)} {D^2}(z)}}{{E(z)D'(z) + 1}} = E(z)D'(z) - 1.
\label{equa:okd}
\end{equation}
Since the left-hand side (lhs)  of Eq.~(\ref{equa:okd}) is a nonzero when $z\neq 0$ if $\Omega_K^{(0)}$ is nonvanishing, the null test of cosmic curvature $\Omega_K^{(0)}$ is equivalent to the null test of the whole lhs of Eq.~(\ref{equa:okd}). If we define
\begin{equation}
{\mathcal O_K(z)}\equiv\frac{{\Omega_K^{(0)} {D^2}(z)}}{{E(z)D'(z) + 1}},
\label{equa:O}
\end{equation}
then a flat universe implies
\begin{equation}
{\mathcal O_K(z)} = E(z)D'(z) - 1 =0
\label{equa:nulltest}
\end{equation}
is always true at any redshift. A deviation from it will indicate a signal of nonvanishing cosmic curvature. Note that the theoretical value of $\mathcal O_K(z)$ is always zero at $z=0$. So the null test is just to check whether there exists  signal of nonvanishing $\mathcal O_K(z)$ at nonzero redshifts.

Thus we should use current observational data sets to reconstruct $E(z)$ and $D'(z)$, independently, and combine these two reconstructions to test whether the relation $E(z)D'(z) - 1 =0$ holds at arbitrary redshifts. We want to stress here that the null test is totally geometrical and cosmological model independent, the reconstructions of $E(z)$ and $D'(z)$ are directly derived from observational data sets.

\section{Null test using $H(z)$  and supernovae data \label{sec:nulltest}}

Given some  observational  data sets, it is crucial to use a model-independent method to reconstruct $E(z)$, $D(z)$, and its derivative $D'(z)$. There are many different methodologies to reconstruct functions from the data.  For a brief analysis, see~\cite{Vitenti:2015aaa}. Since we should use a nonparametric approach to smooth the data and to reconstruct the derivative, the so-called Gaussian processes~\cite{Holsclaw:2010nb,Holsclaw:2010sk,Holsclaw:2011wi,Seikel:2012uu} are very suitable for our purpose.

\subsection{Gaussian processes}

The Gaussian processes allow one to reconstruct a function from data without assuming a parametrization for it. Here, we use the Gaussian processes in Python (GaPP)~\cite{Seikel:2012uu}. This GP code has been used in various papers for different studies ~\cite{Seikel:2012uu,Seikel:2012cs,Bilicki:2012ub,Seikel:2013fda,Yahya:2013xma,Busti:2014dua,Zhang:2014eux,Busti:2015aqa,Cai:2015zoa}. The distribution over functions  provided by GP is suitable to describe the observational  data. At each point $z$, the reconstructed function $f(z)$ is also a Gaussian distribution with a mean value and Gaussian error. The functions at different points $z$ and $\tilde{z}$ are related by a covariance function $k(z,\tilde{z})$, which only depends on a set of hyperparameters $\ell$ and $\sigma_f$. Here, $\ell$ gives a measure of the coherence length of the correlation in the $x$ direction and $\sigma_f$ denotes the overall amplitude of the correlation in the $y$ direction. Both of them will be optimized by GP with the observational data sets.
In contrast to actual parameters, GP does not specify the form of the reconstructed function. Instead, it characterizes the typical changes of the function. The detailed analysis and description of the GP method can be found in~\cite{Seikel:2012uu,Seikel:2013fda}.

\subsection{Hubble rate data and Union 2.1}

Following~\cite{Seikel:2012cs,Bilicki:2012ub}, we proceed to an analysis based on observational Hubble data compiled from several sources, independent of SNeIa. We combine measurements of $H(z)$ obtained with two methods. One is cosmic chronometers, which are mainly passively evolving galaxies. There are $21$ data points compiled by Moresco \emph{et al}.~\cite{Moresco:2012by,Moresco:2015cya}. The other is radial baryon acoustic oscillations from galaxy clustering in redshift surveys, which gives seven data points of Hubble parameters from different experiments~\cite{Gaztanaga:2008xz,Chuang:2011fy,Blake:2012pj,Reid:2012sw}. We summarize the total $28$ data points in Table~\ref{tab:hdata}.

We normalize $H(z)$ using $H_0=70$km/(s Mpc); thus, we get the observational data points of $E(z)$. Then we can use GP method to reconstruct $E(z)$. Note that $H_0$ is just a normalization factor, whose value will not influence our null test Eq.~(\ref{equa:nulltest}).

\begin{table}
\begin{centering}\begin{tabular}{cccc}
\hline
Index & $z$ & $H(z)$ & Refs. \tabularnewline
\hline
1 & $0.090$ & $69\pm 12$ & \cite{Moresco:2012by} \tabularnewline
\hline
2 & $0.170$ & $83\pm 8$ &\cite{Moresco:2012by} \tabularnewline
\hline
3 & $0.179$ & $75\pm 4$ &\cite{Moresco:2012by} \tabularnewline
\hline
4 & $0.199$ & $75\pm 5$ &\cite{Moresco:2012by} \tabularnewline
\hline
5 & $0.240$ & $79.69\pm 2.32$ & \cite{Gaztanaga:2008xz} \tabularnewline
\hline
6 & $0.270$ & $77\pm 14$ &\cite{Moresco:2012by} \tabularnewline
\hline
7 & $0.350$ & $82.1\pm 4.9$ & \cite{Chuang:2011fy} \tabularnewline
\hline
8 & $0.352$ & $83\pm 14$ &\cite{Moresco:2012by} \tabularnewline
\hline
9 & $0.400$ & $95\pm 17$ &\cite{Moresco:2012by} \tabularnewline
\hline
10 & $0.430$ & $86.45\pm 3.27$ & \cite{Gaztanaga:2008xz} \tabularnewline
\hline
11 & $0.440$ & $82.6\pm 7.8$ &  \cite{Blake:2012pj} \tabularnewline
\hline
12 & $0.480$ & $97\pm 62$ & \cite{Moresco:2012by} \tabularnewline
\hline
13 & $0.570$ & $92.4\pm 4.5$ & \cite{Reid:2012sw} \tabularnewline
\hline
14 & $0.593$ & $104\pm 13$ & \cite{Moresco:2012by} \tabularnewline
\hline
15 & $0.600$ & $87.9\pm 6.1$ & \cite{Blake:2012pj} \tabularnewline
\hline
16 & $0.680$ & $92\pm 8$ & \cite{Moresco:2012by} \tabularnewline
\hline
17 & $0.730$ & $97.3\pm 7$ &  \cite{Blake:2012pj} \tabularnewline
\hline
18 & $0.781$ & $105\pm 12$ & \cite{Moresco:2012by} \tabularnewline
\hline
19 & $0.875$ & $125\pm 17$ & \cite{Moresco:2012by} \tabularnewline
\hline
20 & $0.880$ & $90\pm 40$ & \cite{Moresco:2012by} \tabularnewline
\hline
21 & $0.900$ & $117\pm 23$ & \cite{Moresco:2012by} \tabularnewline
\hline
22 & $1.037$ & $154\pm 20$ & \cite{Moresco:2012by} \tabularnewline
\hline
23 & $1.300$ & $168\pm 17$ & \cite{Moresco:2012by} \tabularnewline
\hline
24 & $1.363$ & $160\pm 33.6$ & \cite{Moresco:2015cya} \tabularnewline
\hline
25 & $1.430$ & $177\pm 18$ &\cite{Moresco:2012by} \tabularnewline
\hline
26 & $1.530$ & $140\pm 14$ & \cite{Moresco:2012by} \tabularnewline
\hline
27 & $1.750$ & $202\pm 40$ & \cite{Moresco:2012by} \tabularnewline
\hline
28 & $1.965$ & $186.5\pm 50.4$ & \cite{Moresco:2015cya} \tabularnewline
\hline
\end{tabular}\par\end{centering}
\caption{$H(z)$ measurements from different surveys using passively evolving galaxies and radial BAO.
\label{tab:hdata}}
\end{table}

To reconstruct $D(z)$, we use SNeIa Union 2.1 data sets~\cite{Suzuki:2011hu}, which contain $580$ SNeIa data. We transform the distance modulus $m-M$ given in the data set to $D$ using
\begin{equation}
m-M+5\log\left[\frac{H_0}{c}\right]-25=5\log\left[(1+z)D\right].
\end{equation}
For consistency, here we also use $H_0=70$km/(s Mpc) to normalize $d_L(z)$. Obtaining these $580$ observational data points of $D(z)$, we can also use GP method to reconstruct $D(z)$ and its derivative $D'(z)$. Finally, we combine the reconstructions of $E(z)$ and $D'(z)$ and apply them to the null test of $\mathcal O_K(z)$ in Eq.~(\ref{equa:nulltest}). We stress again that the null test is model independent, so we need not assume any cosmological model, and the two data sets of $H(z)$ and supernovae are also independent of each other.

\begin{figure*}
\includegraphics[width=0.3\textwidth]{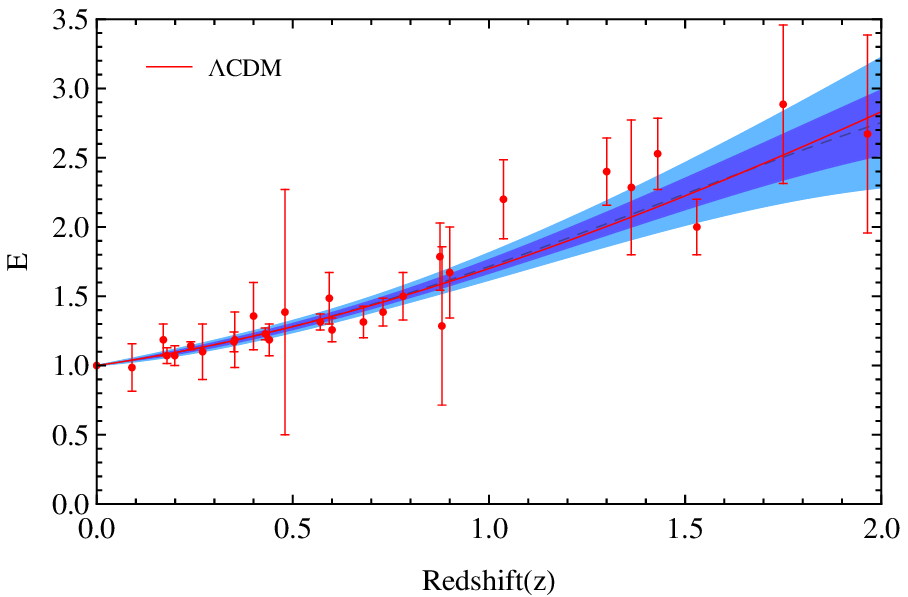}
\includegraphics[width=0.3\textwidth]{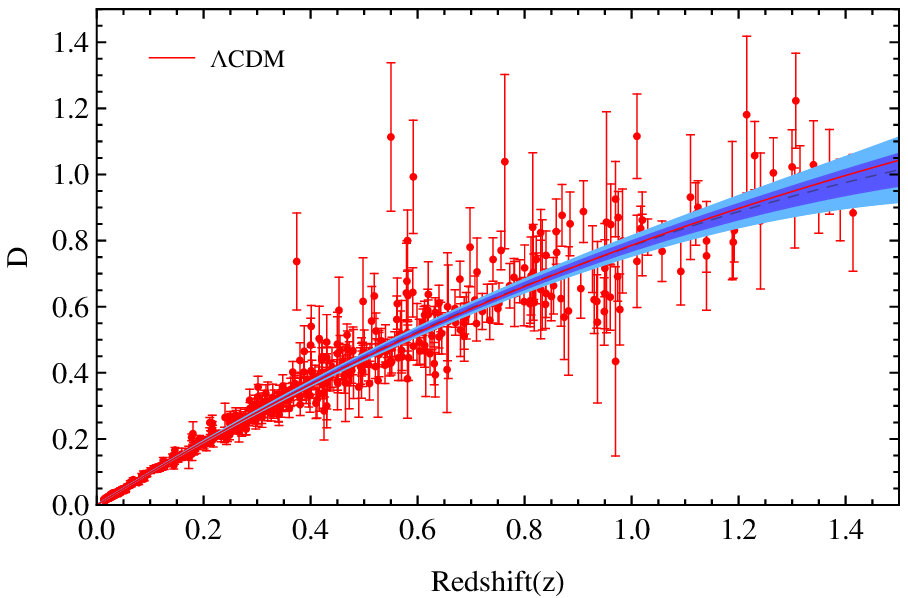}
\includegraphics[width=0.3\textwidth]{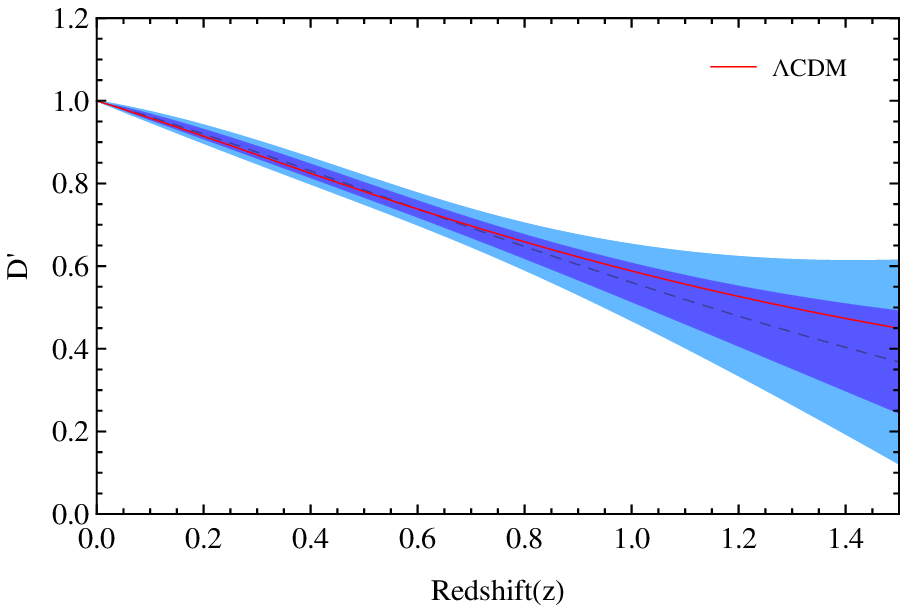}
\caption{Gaussian precess reconstruction of $E(z)$ (left) from CC+BAO, $D(z)$ (middle), and $D'(z)$ (right) from Union 2.1. The shaded blue regions are the $68\%$ and $95\%$ C.L. of the reconstruction. The flat $\Lambda$CDM model (red line) with $\Omega_{m0}=0.27$ is also shown.}
\label{fig:realhD}
\end{figure*}

\begin{figure}
\includegraphics[width=0.4\textwidth]{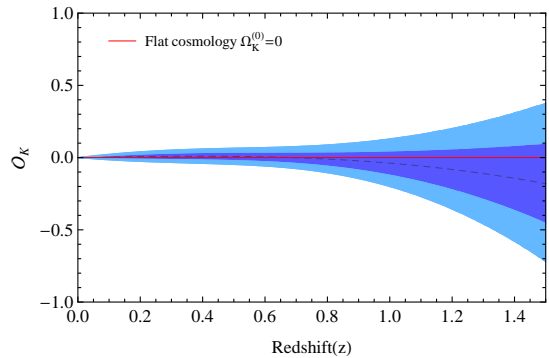}
\caption{Reconstruction of $\mathcal O_K(z)$ from CC+BAO and Union 2.1. The shaded blue regions are the $68\%$ and $95\%$ C.L. of the reconstruction. The red line corresponds to the flat universe $\Omega_K^{(0)}=0$.}
\label{fig:Oreal}
\end{figure}

\subsection{Null test}

Having obtained the total 28 data points of $E(z)$ and 580 points of $D(z)$, we now use the GP method to reconstruct them, respectively.

We can see from Fig.~\ref{fig:realhD} that all of the reconstructions of $E(z)$, $D(z)$, and $D'(z)$ are consistent very well with
the flat $\Lambda$CDM model, which we assume for comparison. The dashed blue line is the mean of the reconstruction and the shaded blue regions are the $68\%$ and $95\%$ confidence level (C.L.) of the reconstruction. As expected, at higher redshifts, the errors become large due to the poor quality data in that region. Using the reconstructions of $E(z)$ and $D'(z)$, we apply Monte Carlo sampling to determine the $\mathcal O_K(z)$ in Eq.~(\ref{equa:nulltest}) at each point $z$, which we want to reconstruct. In Fig.~\ref{fig:Oreal}, it is shown that the reconstructed $\mathcal O_K(z)$  is consistent with the vanishing cosmic curvature, falling in the $1\sigma$ limit. It tells us that there is no significant signal to indicate the deviation of the cosmic curvature $\Omega_K^{(0)}$ from 0 at the current observational data [$H(z)$ and supernovae] level. In addition, let us mention that the mean value of the cosmic curvature is negative in the high
redshift region, which is also consistent with the results from model-dependent constraints in the literature.

\begin{figure*}
\includegraphics[width=0.35\textwidth]{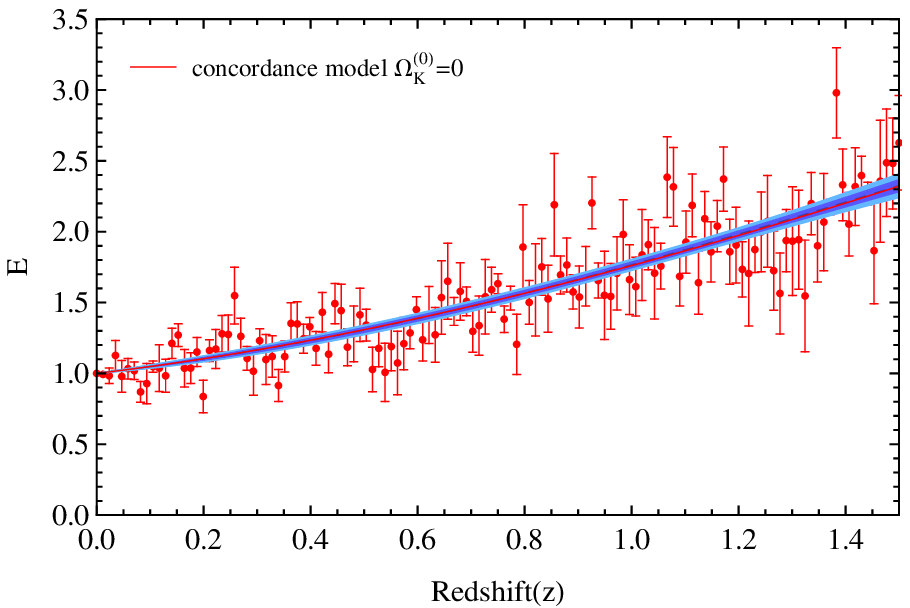}\quad
\includegraphics[width=0.35\textwidth]{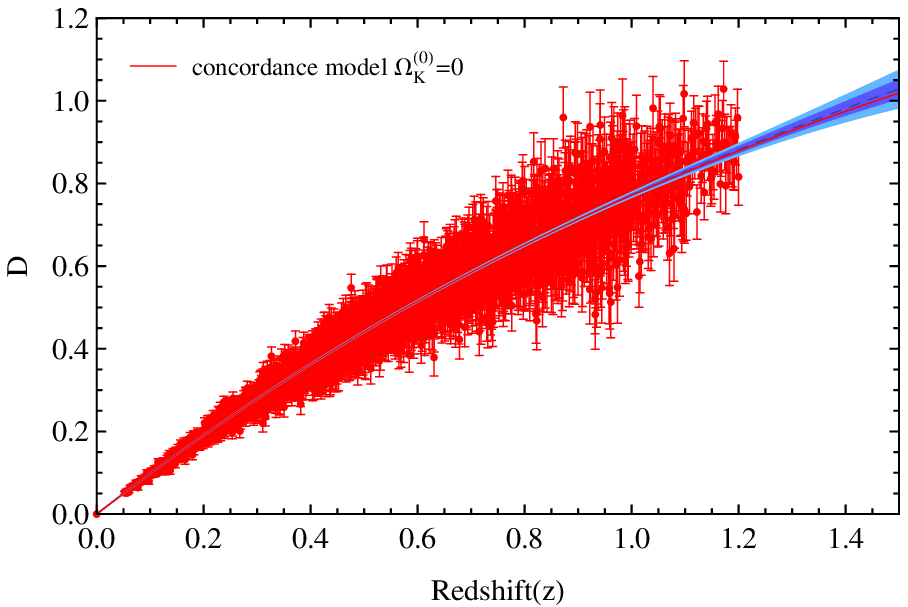}\\
\includegraphics[width=0.35\textwidth]{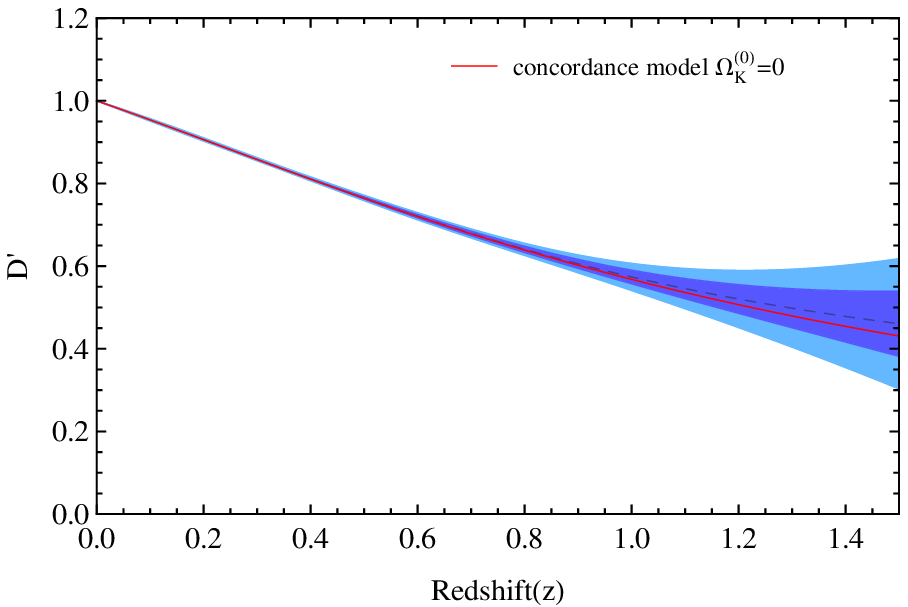}\quad
\includegraphics[width=0.35\textwidth]{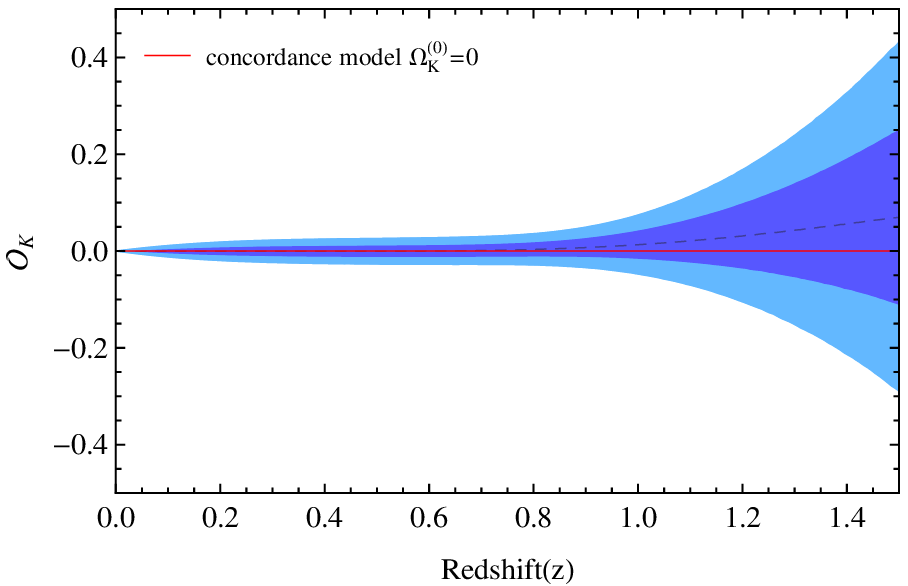}
\caption{Gaussian process reconstruction of $E(z)$, $D(z)$ ({\em top}), $D'(z)$, and reconstruction of $\mathcal O_K(z)$({\em bottom}) obtained from a mock data set of $E(z)$ and future DES, assuming the concordance model with $\Omega_K^{(0)}=0$ (red line). The dashed blue line is the mean of the reconstruction, and the shaded blue regions are the $68\%$ and $95\%$ C.L. of the reconstruction, respectively.}
\label{fig:mockOLCDM}
\end{figure*}

\begin{figure*}
\includegraphics[width=0.35\textwidth]{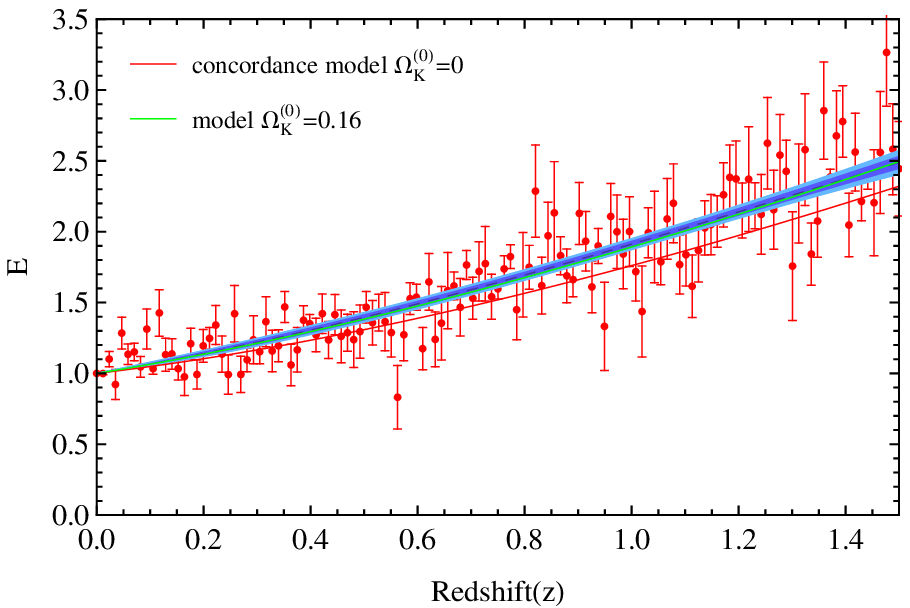}\quad
\includegraphics[width=0.35\textwidth]{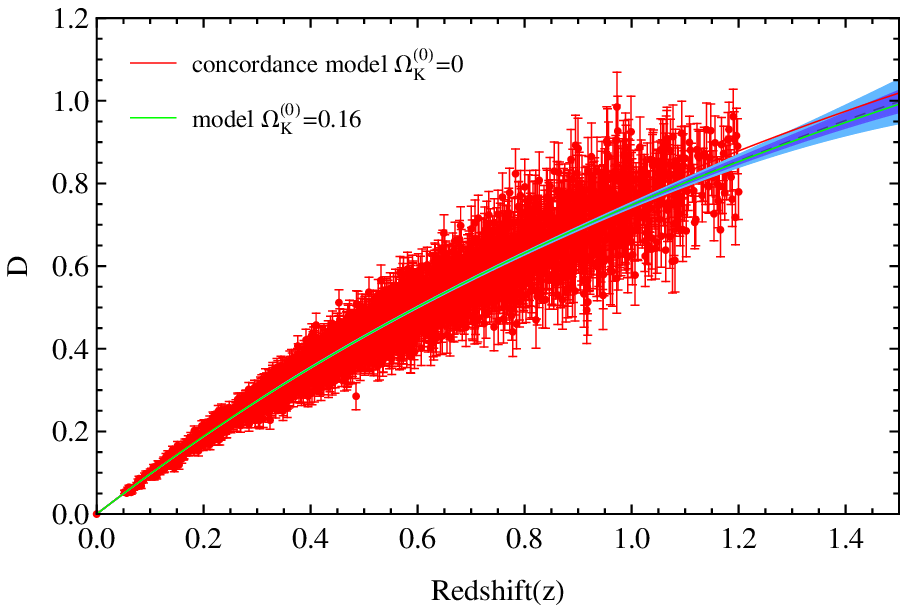}\\
\includegraphics[width=0.35\textwidth]{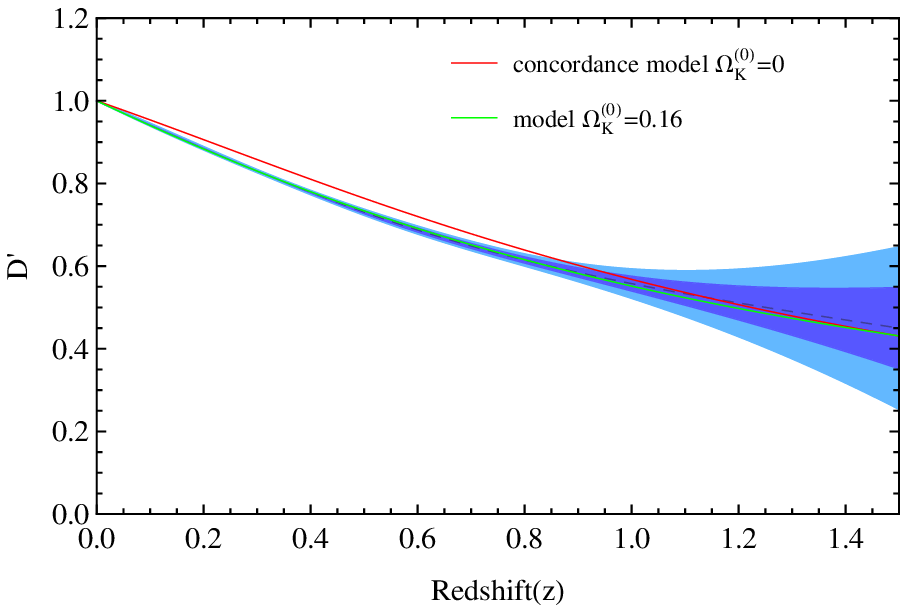}\quad
\includegraphics[width=0.35\textwidth]{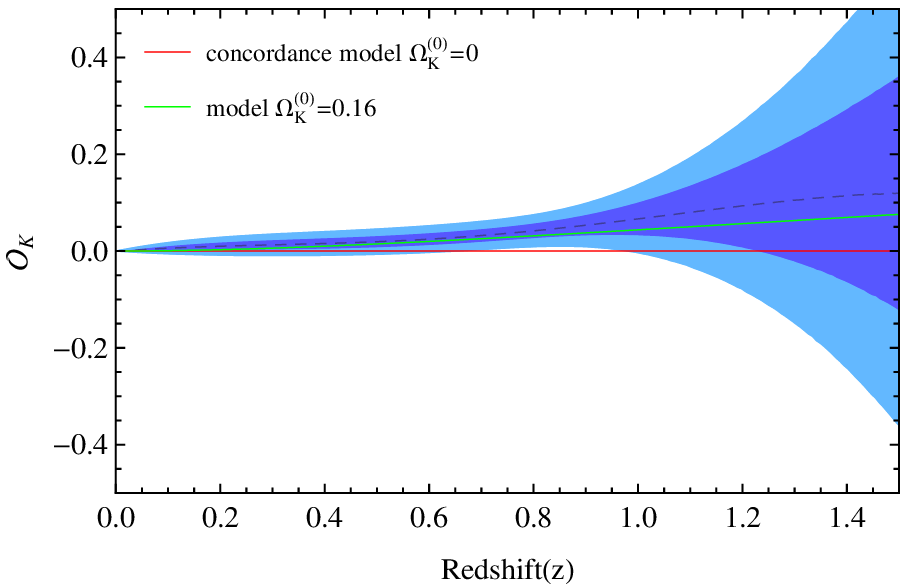}
\caption{Gaussian process reconstruction of $E(z)$, $D(z)$ ({\em top}), $D'(z)$, and reconstruction of $\mathcal O_K(z)$({\em bottom}) obtained from a mock data set of $E(z)$ and future DES, assuming the fiducial model with $\Omega_K^{(0)}=0.16$ (green line). The dashed blue line is the mean of the reconstruction, and the shaded blue regions are the $68\%$ and $95\%$ C.L. of the reconstruction, respectively. The concordance model $\Omega_K^{(0)}=0$ is also shown (red line).}
\label{fig:mockO16}
\end{figure*}

\begin{figure*}
\includegraphics[width=0.35\textwidth]{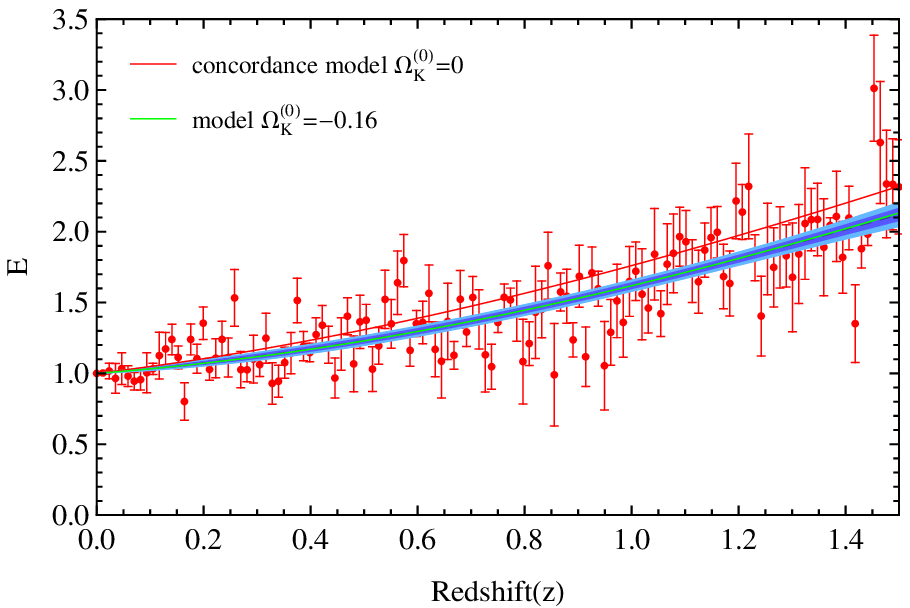}\quad
\includegraphics[width=0.35\textwidth]{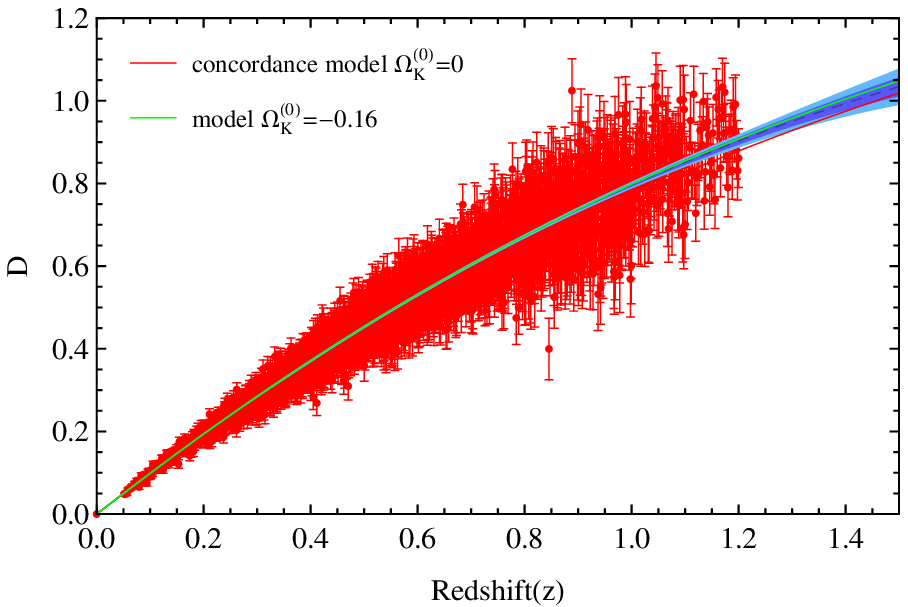}\\
\includegraphics[width=0.35\textwidth]{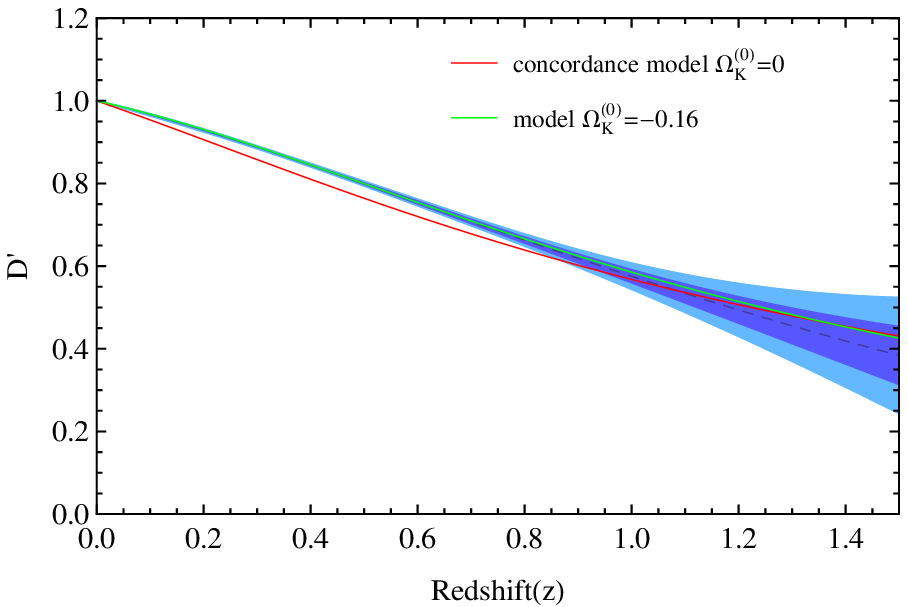}\quad
\includegraphics[width=0.35\textwidth]{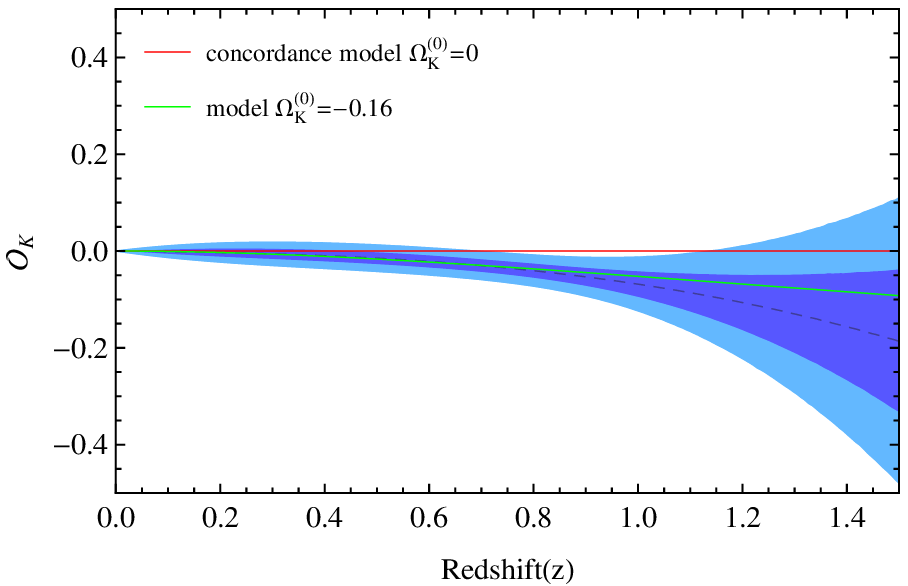}
\caption{Gaussian process reconstruction of $E(z)$, $D(z)$ ({\em top}), $D'(z)$, and reconstruction of $\mathcal O_K(z)$({\em bottom}) obtained from a mock data set of $E(z)$ and future DES, assuming the fiducial model with $\Omega_K^{(0)}=-0.16$ (green line). The dashed blue line is the mean of the reconstruction, and the shaded blue regions are the $68\%$ and $95\%$ C.L. of the reconstruction, respectively. The concordance model $\Omega_K^{(0)}=0$ is also shown (red line).}
\label{fig:mockOm16}
\end{figure*}

\subsection{Mock data}

To demonstrate how a large number of data with what accuracy of the error will affect our null test when reconstructing $E(z)$ and $D(z)$, we firstly simulate a data set of 128 points for $E(z)$. Adopting the methodology in~\cite{Ma:2010mr}, we draw the error from a Gaussian distribution: $\sigma_{E} \sim \mathcal{N}(\bar{\sigma},\epsilon)$ with $\bar{\sigma} =(\sigma_+ + \sigma_- )/2$ and $\epsilon = (\sigma_+ - \sigma_- )/4$, where $\sigma_+$ and $\sigma_-$ are the two straight lines that bound the uncertainties $\sigma(z)$ of the observational $E(z)$ data from above and below, respectively. Then $E(z)_{sim}$ is sampled from the Gaussian distribution $E(z)_{sim} \sim \mathcal{N}(E(z)_{fid},\sigma_{E})$, where $E(z)_{fid}$ is the theoretical value from the fiducial model.

As for simulated $D(z)$ data, we create mock data sets of future SNeIa according to the Dark Energy Survey (DES)~\cite{Bernstein:2011zf}. The DES is expected to obtain high quality light curves for about 4000 SNeIas from $z=0.05$ to $z=1.2$. From Table $14$ in~\cite{Bernstein:2011zf}, we can calculate the errors of $D$, $\sigma_{D}$, and the corresponding numbers of SNeIa for each redshift bin. At every redshift point $z$, $D(z)_{sim}$ is sampled from the normal distribution $D(z)_{sim} \sim \mathcal{N}(D(z)_{fid},\sigma_{D})$.

Having obtained the simulated data sets of $E(z)$ and $D(z)$, we use GP method to reconstruct $E(z)$, $D(z)$, and $D'(z)$. Then we combine the reconstructions of $E(z)$ and $D'(z)$, using Monte Carlo sampling to determine the $\mathcal O_K(z)$ and check whether GP can recover its theoretical value and distinguish it from $\Omega_K^{(0)}=0$.

We now simulate the data points for three different fiducial models: concordance model, namely, $\Lambda$CDM model with $\Omega_K^{(0)} =0$ and $\Omega_m=0.3$; two models with nonvanishing cosmic curvature, $\Omega_K^{(0)} =\pm0.16$ and $\Omega_m=0.3$. We want to check whether the GP method can detect or recover all of them and distinguish them from each other. The results are shown in Figs.~\ref{fig:mockOLCDM}-\ref{fig:mockOm16}, respectively.

\begin{figure*}
\subfloat[]{
\includegraphics[width=0.4\textwidth]{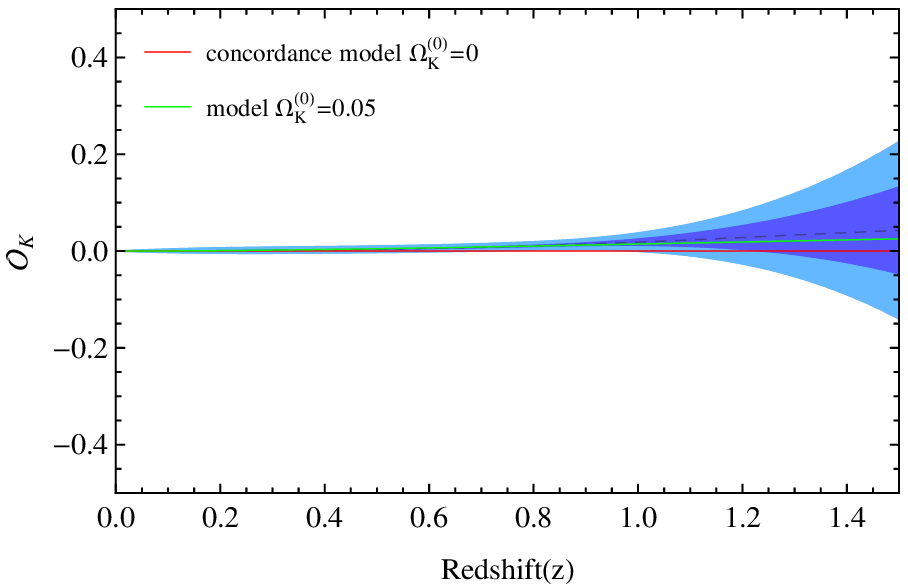}}\quad
\subfloat[]{
\includegraphics[width=0.4\textwidth]{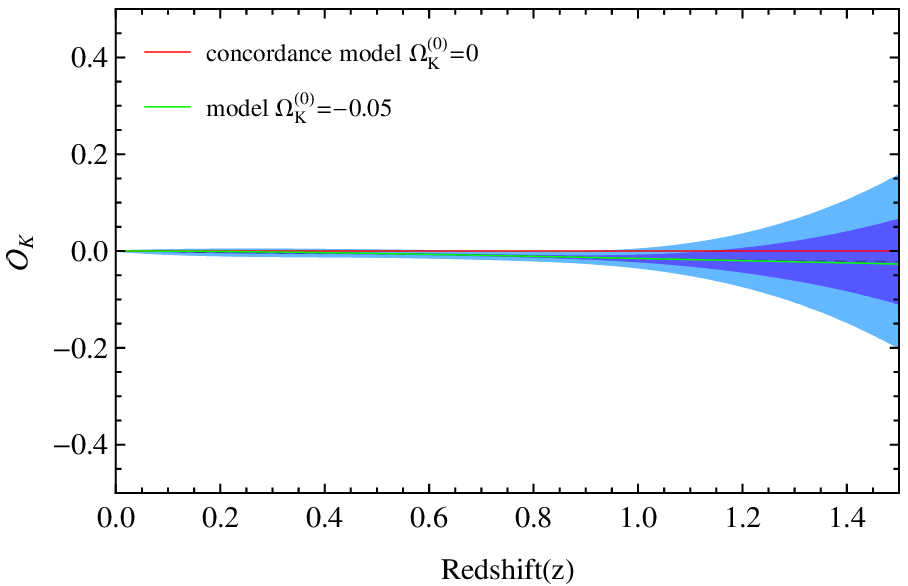}}\\
\subfloat[]{
\includegraphics[width=0.4\textwidth]{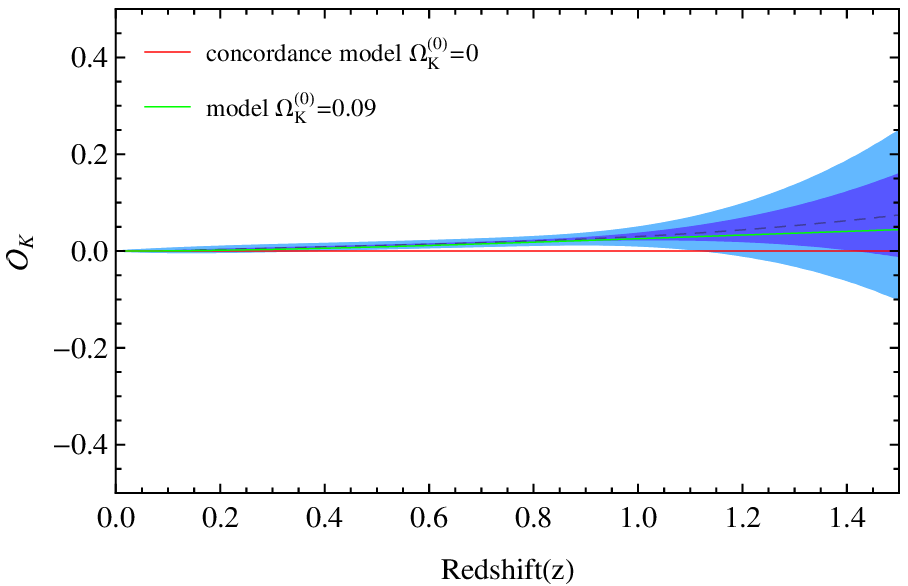}}\quad
\subfloat[]{
\includegraphics[width=0.4\textwidth]{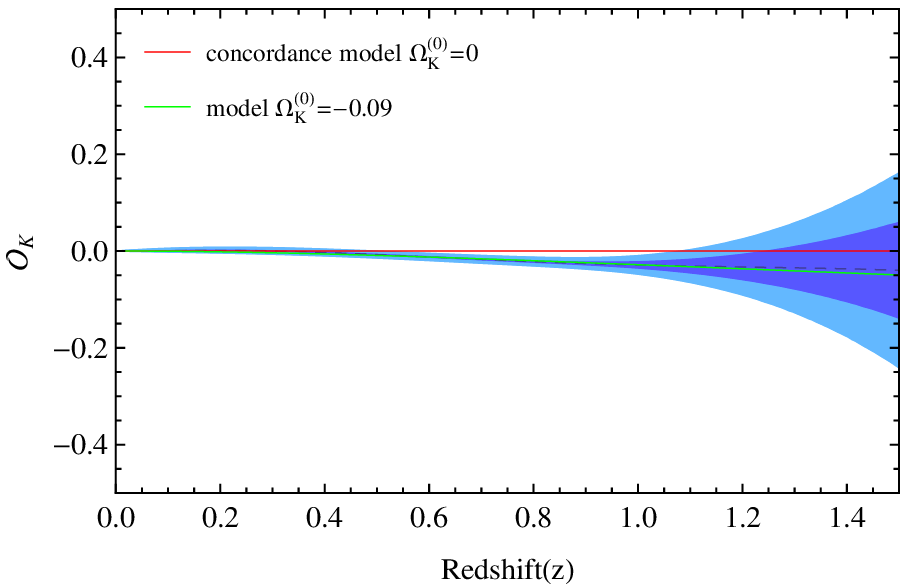}}\\
\subfloat[]{
\includegraphics[width=0.4\textwidth]{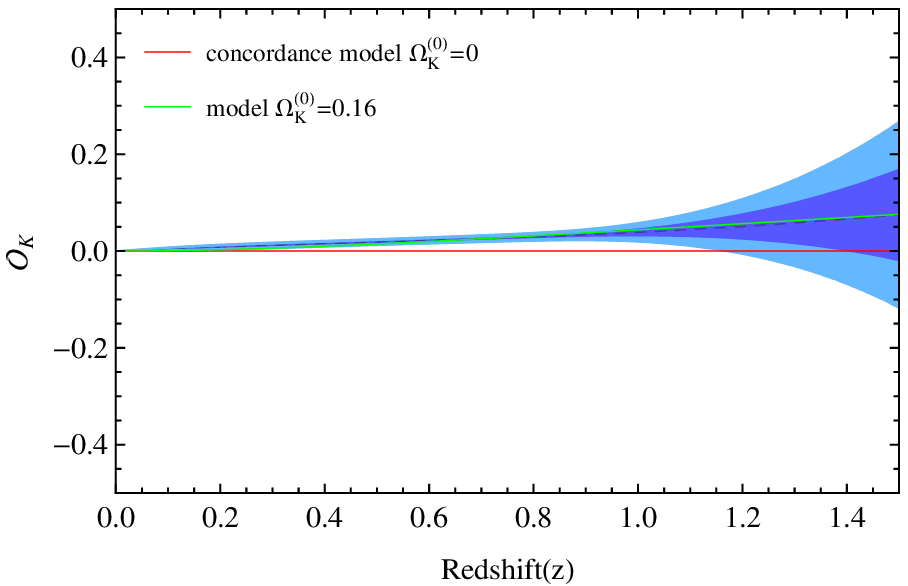}}\quad
\subfloat[]{
\includegraphics[width=0.4\textwidth]{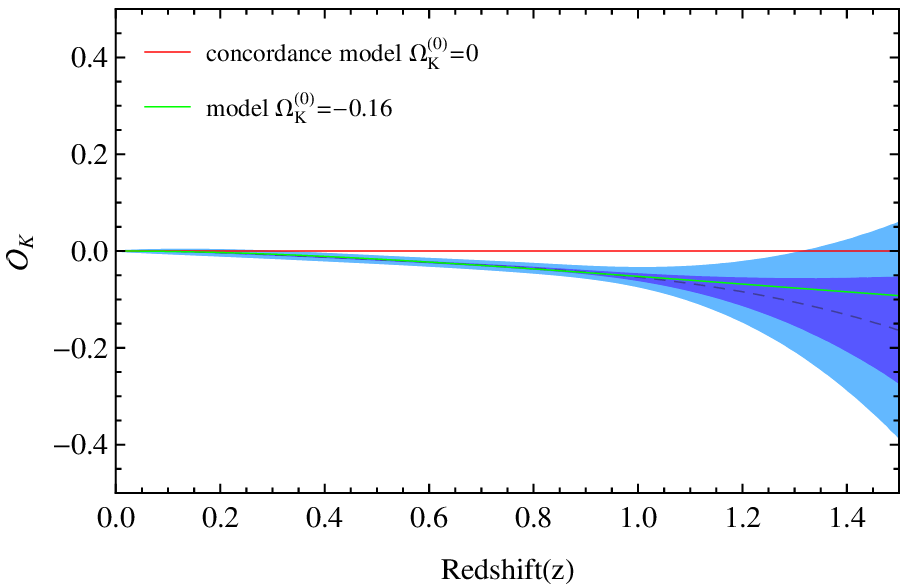}}
\caption{Reconstructions of $\mathcal O_K(z)$ for models with different $\Omega_K^{(0)}$, the errors are a quarter of the original errors of the mock data sets. (a) $\Omega_K^{(0)}=0.05$, (b) $\Omega_K^{(0)}=-0.05$, (c) $\Omega_K^{(0)}=0.09$, (d) $\Omega_K^{(0)}=-0.09$, (e) $\Omega_K^{(0)}=0.16$, (f) $\Omega_K^{(0)}=-0.16$, The shaded blue regions are the $68\%$ and $95\%$ C.L. for the reconstruction.}
\label{fig:mockse}
\end{figure*}

We can see from Fig.~\ref{fig:mockOLCDM} that $E(z)$, $D(z)$, and $D'(z)$ are all reconstructed very well from the mock data sets assuming the concordance model $\Omega_K^{(0)}=0$. And the reconstructed $\mathcal O_K(z)$ is also consistent with the concordance model nicely. As expected, at higher redshifts, the errors become large  due to the poor quality data in that region. Furthermore, we see from Figs.~\ref{fig:mockO16} and~\ref{fig:mockOm16} that the reconstructions of model $\Omega_K^{(0)}=\pm0.16$ also recover the fiducial model very well, falling in the $1\sigma$ limit and obviously deviating from the concordance model $\Omega_K^{(0)}=0$ at $95\%$ C.L. The large errors and not so good reconstructions at high redshifts ($z>1.0$) are due to the poor quality data in that region for there are no simulated data of $D(z)$ at $z>1.2$. Anyway, we can claim that with these quantities and errors of the observational data for $E(z)$ and $D(z)$, the GP method at least has the ability to detect the cosmic curvature $\Omega_K^{(0)}\geq0.16$  and to rule out $\Omega_K^{(0)}=0$ at $2\sigma$ C.L.

However, as we have pointed out that the theoretical value of $\mathcal O_K(z)$ is always zero at $z=0$. So the theoretical values of $\mathcal O_K(z)$ for these models with different $\Omega_K^{(0)}$ deviates from the model with $\Omega_K^{(0)}=0$ tiny at low redshifts. As $z$ increases, the difference becomes large. As a result, we can see from Figs.~\ref{fig:mockO16} and~\ref{fig:mockOm16} that it is very hard to rule out $\Omega_K^{(0)}=0$ at low redshifts. Maybe it can work out at a middle redshift ($0.6<z<1.0$) as Figs.~\ref{fig:mockO16} and~\ref{fig:mockOm16} show, but it is not always helpful when  $\Omega_K^{(0)}$ is smaller. So, if we want to detect a model with $\Omega_K^{(0)}<0.16$ or even smaller and can rule out $\Omega_K^{(0)}=0$ at $95\%$ C.L., a larger number and higher quality data sets are required.

Figure~\ref{fig:mockse} shows the reconstructed $\mathcal O_K(z)$ also using the mock data sets of $E(z)$ and $D(z)$ but with a quarter of the errors.  The reconstructions are more nice. And the null test is more precise; it can detect the model with $\Omega_K^{(0)}=0.05$ or even smaller, ruling out  $\Omega_K^{(0)}=0$ at $95\%$ C.L.

\section{ discussions and conclusions \label{sec:discussion}}

In this paper, we have introduced a nonparametric approach to making a null test of the cosmic curvature. Using the Gaussian process method, we  reconstructed the Hubble rate and distance-redshift relation [$E(z)$, $D(z)$, and $D'(z)$], independently. In the
reconstruction, we needed not to assume any cosmological model.  By combining the reconstructions of $E(z)$ and $D'(z)$, we can determine $\mathcal O_K(z)$, which is related to the cosmic curvature $\Omega_K^{(0)}$. We have shown that $\Omega_K^{(0)}=0$ is consistent with current data sets (CC+BAO, Union 2.1), falling within the $1\sigma$ limit, although the mean of the reconstructed curvature ${\cal O}_K$ is negative in the high redshift region, which is also consistent with the results from the model-dependent constraints in the literature. In addition, note that the mean of the reconstructed curvature ${\cal O}_K$ in Fig.~\ref{fig:mockOLCDM} is positive in the high redshift region for the fiducial flat $\Lambda$CDM model. In this sense, our result shown in Fig.\ref{fig:Oreal} indicates that there is a little possibility for a closed universe.

To demonstrate how a large number of data sets with different accuracies and errors will affect our null test, we create mock data sets of $E(z)$ and $D(z)$ using the methodology proposed in Refs.~\cite{Ma:2010mr} and~\cite{Bernstein:2011zf}. We found that with current quality of simulated data, the GP method can recover and distinguish models with a different cosmic curvature with an order of $10^{-1}$ from $\Omega_K^{(0)}=0$. However, if we want to detect even smaller $\Omega_K^{(0)}$, we should decrease the uncertainties of the data sets. Making the errors a quarter of the mock data, it can rule out $\Omega_K^{(0)}=0$  from the  cosmic curvature with an order of $10^{-2} $ at $95\%$ C.L.

Based on a one-parameter extension to the six-parameter $\Lambda$CDM model,
CMB data~\cite{Hinshaw:2012aka,Ade:2013zuv,Ade:2015xua} provide a strong constraint on $\Omega_K^{(0)}$.
This constraint can be improved dramatically by adding BAO data that break the geometric degeneracy between $\Omega_K^{(0)}$ and $H_0$.
However, the model assumption and priors on model parameters may bias estimates of the cosmic curvature.
Here, we proposed a model-independent method for reconstructing the function ${\cal O}_K(z)$ with redshift,
which characterizes a deviation from a flat universe.
As we can see from Eq.~\eqref{equa:nulltest}, ${\cal O}_K(z)$ is completely and directly determined by the Hubble parameter and luminosity distance.
If we can reconstruct the Hubble parameter and luminosity distance, the reconstruction of the cosmic curvature is straightforward.
GP is suitable for our purpose because it can smooth data and reconstruct a function model independently.
Therefore, we can test the cosmic curvature neither assuming models nor imposing priors.
Although current low-redshift data put weaker constraints on the cosmic curvature than CMB data,
this method provides a new cross-check of the cosmic curvature using future data from large-scale structure measurements.


\begin{acknowledgements}
This work is supported by the Strategic Priority Research Program of the Chinese Academy of Sciences, Grant No.XDB09000000.
Z.K.G is supported by the National Natural Science Foundation of China Grants No. 11575272 and No. 11335012.
\end{acknowledgements}



\begin{thebibliography}{99}

\bibitem{Ichikawa:2005nb}
  K.~Ichikawa and T.~Takahashi,
  Phys.\ Rev.\ D {\bf 73}, 083526 (2006)
  [astro-ph/0511821].

\bibitem{Ichikawa:2006qb}
  K.~Ichikawa, M.~Kawasaki, T.~Sekiguchi and T.~Takahashi,
  JCAP {\bf 0612}, 005 (2006)
  [astro-ph/0605481].

\bibitem{Clarkson:2007bc}
  C.~Clarkson, M.~Cortes and B.~A.~Bassett,
  JCAP {\bf 0708}, 011 (2007)
  [astro-ph/0702670 [ASTRO-PH]].

\bibitem{Gong:2006gs}
  Y.~G.~Gong and A.~Wang,
  Phys.\ Rev.\ D {\bf 75}, 043520 (2007)
  [astro-ph/0612196].

\bibitem{Eisenstein:2005su}
  D.~J.~Eisenstein {\it et al.} [SDSS Collaboration],
  Astrophys.\ J.\  {\bf 633}, 560 (2005)
  [astro-ph/0501171].

\bibitem{Tegmark:2006az}
  M.~Tegmark {\it et al.} [SDSS Collaboration],
  Phys.\ Rev.\ D {\bf 74}, 123507 (2006)
  [astro-ph/0608632].

\bibitem{Zhao:2006qg}
  G.~B.~Zhao, J.~Q.~Xia, H.~Li, C.~Tao, J.~M.~Virey, Z.~H.~Zhu and X.~Zhang,
  Phys.\ Lett.\ B {\bf 648}, 8 (2007)
  [astro-ph/0612728].

\bibitem{Wright:2007vr}
  E.~L.~Wright,
  Astrophys.\ J.\  {\bf 664}, 633 (2007)
  [astro-ph/0701584].

\bibitem{Ade:2015xua}
  P.~A.~R.~Ade {\it et al.}  [Planck Collaboration],
  arXiv:1502.01589 [astro-ph.CO].

\bibitem{Shafieloo:2009hi}
  A.~Shafieloo and C.~Clarkson,
  Phys.\ Rev.\ D {\bf 81}, 083537 (2010)
  [arXiv:0911.4858 [astro-ph.CO]].

\bibitem{Li:2014yza}
  Y.~L.~Li, S.~Y.~Li, T.~J.~Zhang and T.~P.~Li,
  Astrophys.\ J.\  {\bf 789}, L15 (2014)
  [arXiv:1404.0773 [astro-ph.CO]].

\bibitem{Vitenti:2015aaa}
  S.~D.~P.~Vitenti and M.~Penna-Lima,
  JCAP {\bf 1509}, no. 09, 045 (2015)
  [arXiv:1505.01883 [astro-ph.CO]].

\bibitem{Holsclaw:2010nb}
  T.~Holsclaw, U.~Alam, B.~Sanso, H.~Lee, K.~Heitmann, S.~Habib and D.~Higdon,
  Phys.\ Rev.\ D {\bf 82}, 103502 (2010)
  [arXiv:1009.5443 [astro-ph.CO]].

\bibitem{Holsclaw:2010sk}
  T.~Holsclaw, U.~Alam, B.~Sanso, H.~Lee, K.~Heitmann, S.~Habib and D.~Higdon,
  Phys.\ Rev.\ Lett.\  {\bf 105}, 241302 (2010)
  [arXiv:1011.3079 [astro-ph.CO]].

\bibitem{Holsclaw:2011wi}
  T.~Holsclaw, U.~Alam, B.~Sanso, H.~Lee, K.~Heitmann, S.~Habib and D.~Higdon,
  Phys.\ Rev.\ D {\bf 84}, 083501 (2011)
  [arXiv:1104.2041 [astro-ph.CO]].

\bibitem{Seikel:2012uu}
  M.~Seikel, C.~Clarkson and M.~Smith,
  JCAP {\bf 1206}, 036 (2012)
  [arXiv:1204.2832 [astro-ph.CO]].

\bibitem{Seikel:2012cs}
  M.~Seikel, S.~Yahya, R.~Maartens and C.~Clarkson,
  Phys.\ Rev.\ D {\bf 86}, 083001 (2012)
  [arXiv:1205.3431 [astro-ph.CO]].

\bibitem{Bilicki:2012ub}
  M.~Bilicki and M.~Seikel,
  Mon.\ Not.\ Roy.\ Astron.\ Soc.\  {\bf 425}, 1664 (2012)
  [arXiv:1206.5130 [astro-ph.CO]].

\bibitem{Seikel:2013fda}
  M.~Seikel and C.~Clarkson,
  arXiv:1311.6678 [astro-ph.CO].

\bibitem{Yahya:2013xma}
  S.~Yahya, M.~Seikel, C.~Clarkson, R.~Maartens and M.~Smith,
  Phys.\ Rev.\ D {\bf 89}, no. 2, 023503 (2014)
  [arXiv:1308.4099 [astro-ph.CO]].

\bibitem{Busti:2014dua}
  V.~C.~Busti, C.~Clarkson and M.~Seikel,
  Mon.\ Not.\ Roy.\ Astron.\ Soc.\  {\bf 441}, 11 (2014)
  [arXiv:1402.5429 [astro-ph.CO]].

\bibitem{Zhang:2014eux}
  Y.~Zhang,
  arXiv:1408.3897 [astro-ph.CO].

\bibitem{Busti:2015aqa}
  V.~C.~Busti and C.~Clarkson,
  arXiv:1505.01821 [astro-ph.CO].

\bibitem{Cai:2015zoa}
  T.~Yang, Z.~K.~Guo and R.~G.~Cai,
  Phys.\ Rev.\ D {\bf 91}, no. 12, 123533 (2015)
  [arXiv:1505.04443 [astro-ph.CO]].

\bibitem{Moresco:2012by}
  M.~Moresco, L.~Verde, L.~Pozzetti, R.~Jimenez and A.~Cimatti,
  JCAP {\bf 1207}, 053 (2012)
  [arXiv:1201.6658 [astro-ph.CO]].

\bibitem{Moresco:2015cya}
  M.~Moresco,
  Mon.\ Not.\ Roy.\ Astron.\ Soc.\  {\bf 450}, no. 1, L16 (2015)
  [arXiv:1503.01116 [astro-ph.CO]].


\bibitem{Gaztanaga:2008xz}
  E.~Gaztanaga, A.~Cabre and L.~Hui,
  Mon.\ Not.\ Roy.\ Astron.\ Soc.\  {\bf 399}, 1663 (2009)
  [arXiv:0807.3551 [astro-ph]].

\bibitem{Chuang:2011fy}
  C.~H.~Chuang and Y.~Wang,
  Mon.\ Not.\ Roy.\ Astron.\ Soc.\  {\bf 426}, 226 (2012)
  [arXiv:1102.2251 [astro-ph.CO]].

\bibitem{Blake:2012pj}
  C.~Blake {\it et al.},
  Mon.\ Not.\ Roy.\ Astron.\ Soc.\  {\bf 425}, 405 (2012)
  [arXiv:1204.3674 [astro-ph.CO]].

\bibitem{Reid:2012sw}
  B.~A.~Reid {\it et al.},
  Mon.\ Not.\ Roy.\ Astron.\ Soc.\  {\bf 426}, 2719 (2012)
  [arXiv:1203.6641 [astro-ph.CO]].

\bibitem{Suzuki:2011hu}
  N.~Suzuki {\it et al.},
  Astrophys.\ J.\  {\bf 746}, 85 (2012)
  [arXiv:1105.3470 [astro-ph.CO]].

\bibitem{Ma:2010mr}
  C.~Ma and T.~J.~Zhang,
  Astrophys.\ J.\  {\bf 730}, 74 (2011)
  [arXiv:1007.3787 [astro-ph.CO]].

\bibitem{Bernstein:2011zf}
  J.~P.~Bernstein {\it et al.},
  Astrophys.\ J.\  {\bf 753}, 152 (2012)
  [arXiv:1111.1969 [astro-ph.CO]].

\bibitem{Hinshaw:2012aka}
  G.~Hinshaw {\it et al.} [WMAP Collaboration],
  Astrophys.\ J.\ Suppl.\  {\bf 208}, 19 (2013)
  [arXiv:1212.5226 [astro-ph.CO]].

\bibitem{Ade:2013zuv}
  P.~A.~R.~Ade {\it et al.} [Planck Collaboration],
  Astron.\ Astrophys.\  {\bf 571}, A16 (2014)
  [arXiv:1303.5076 [astro-ph.CO]].


\end{thebibliography}
\end{document}